\documentclass[showpacs,preprintnumbers,amsmath,amssymb]{revtex4}
\usepackage{amsmath,amssymb,graphics,epsfig,subfigure}
\usepackage{color}
\usepackage{multirow}
\usepackage{graphicx}
\usepackage{epstopdf}
\usepackage{dcolumn}
\usepackage{amsmath, mathrsfs}
\usepackage{amsmath}
\usepackage{amssymb}
\usepackage{bm}
\usepackage{color}
\usepackage{multirow}
\usepackage{float}
\usepackage[dvipsnames]{xcolor}
\usepackage{hyperref}

\begin{document}
\newcommand {\nn} {\nonumber}
\renewcommand{\baselinestretch}{1.3}

 \baselineskip=0.8cm
\title{Near-horizon polarized images of a rotating hairy Horndeski black hole}

\author{Chengjia Chen$^{1}$, Qiyuan Pan$^{1,2}$\footnote{panqiyuan@hunnu.edu.cn}, and Jiliang Jing$^{1}$\footnote{jljing@hunnu.edu.cn}}

\affiliation{$^1$Department of Physics, Institute of Interdisciplinary Studies, Key Laboratory of Low Dimensional Quantum Structures
    and Quantum Control of Ministry of Education, Synergetic Innovation Center for Quantum Effects and Applications, and Hunan Research Center of the Basic Discipline for Quantum Effects and Quantum Technologies, Hunan Normal University,  Changsha, Hunan 410081, People's Republic of China
    \\
    $ ^2$Center for Gravitation and Cosmology, College of Physical Science and Technology, Yangzhou University, Yangzhou 225009, People's Republic of China}

\begin{abstract}
 \baselineskip=0.6cm
 Recently, Hou \emph{et al.} [Astrophys. J. Lett. \textbf{988}, L51 (2025)] revealed that the Electric Vector Position Angle (EVPA) of polarization vectors in the near-horizon images is governed solely by the spacetime geometry and is irrespective of the plasma flows. Here, we generalize the study to the scenario of a rotating hair black hole within the Horndeski gravity and probe the effects of the hairy parameter on the EVPA.  For a fixed inclination, the hairy parameter enhances the observed EVPA in the slowly rotating case, but decreases it in the rapidly rotating case.  For a fixed black hole spin, the influence of the hairy parameter on the observed EVPA under different observer inclinations is further modulated by the azimuthal angle of the observed polarization vector. The hairy parameter's impact is more distinct in the low inclination case as the azimuthal angle lies within a specific range, but is almost independent of the observer inclination as the azimuthal angle is beyond this specific range. Furthermore, the dependence of the hairy parameter’s impact on the EVPA is stronger with respect to the black hole spin than to the inclination angle. These results could help to further understand the near-horizon polarized images and Horndeski gravity.
\end{abstract}


\maketitle

\newpage
\section{Introduction}
\label{secIntroduction}

 The images of supermassive black holes released by the Event Horizon Telescope (EHT) collaboration \cite{EventHorizonTelescope:2019dse,EventHorizonTelescope:2019pgp,EventHorizonTelescope:2019ggy,EventHorizonTelescope:2021srq,EventHorizonTelescope:2022wkp,EventHorizonTelescope:2022urf} provide compelling evidence for the existence of black holes. The polarization patterns stored in these images  \cite{EventHorizonTelescope:2021bee,EHT:2024nwx,EHT:2024ehx} reveal that there exists the magnetic field around supermassive black holes. Therefore, the polarized images of black holes are expected as novel avenues to explore the matter distribution surrounding black holes, to identify black hole parameters, as well as to test the theory of general relativity. It is well known that the observed polarization images heavily depend on the plasma accretion flows around black holes. Therefore, in order to  accurately extract the information on black holes, it is crucial to disentangle the corresponding plasma effects from the images. Recently, the near-horizon polarization pattern is found to be determined solely by the spacetime geometry and observer's inclination,  while is irrespective of the plasma flows \cite{Hou:2024qqo}. This means that the near-horizon polarization pattern provides a clean and precise probe of extracting the black hole information including the black hole spin and other fundamental parameters. Furthermore, their analysis indicates that such polarization pattern may be detectable with future high-resolution observations, such as EHT arrays and spacebased VLBI. However, their research focuses on only the Kerr and Kerr-Newman-Nut black hole spacetimes. Thus, it is necessary to study near-horizon polarization patterns for other black holes within theories of the alternative gravity.
 
The  simplest and non-trivial modifications of general relativity are probably scalar-tensor theories, which contain a scalar degree of freedom besides the gravitational metric. The Horndeski gravity is one of  the most famous scalar-tensor theories, which plays a significant role in explaining the accelerating expansion of current universe \cite{Kase:2018aps}. Although the Horndeski theory contains higher derivatives of a scalar field, it owns the diffeomorphism invariance and the second-order differential field equations. Another good behavior of Horndeski gravity is that there is no the Ostrogradsky instability \cite{gw,Deffayet:2011gz}. Therefore, the Horndeski theory provides a natural platform for verifying the no-hair theorem of black holes. The hairy black holes in the Horndeski gravity have been widely studied by including the radially dependent
scalar field \cite{Rinaldi:2012vy,Cisterna:2014nua,Feng:2015wvb,Sotiriou:2013qea,Miao:2016aol,Kuang:2016edj,Babichev:2016rlq,Benkel:2016rlz,Filios:2018xvy,Cisterna:2018hzf,Giusti:2021sku,Cisterna:2015uya,Anabalon:2013oea,Fontana:2018fof} and the time-dependent scalar field \cite{Babichev:2013cya,Babichev:2017lmw,BenAchour:2018dap,Takahashi:2019oxz,Minamitsuji:2019shy,Arkani-Hamed:2003juy}. Recently, a static and
spherical symmetric hairy black hole solution of Horndeski
gravity has been obtained with the quartic scalar field model \cite{Bergliaffa:2021diw}. This Horndeski hairy carries a hairy parameter with the dimension of length.
The presence of the hairy parameter yields that the metric  contains an additional log-term deforming from the standard Schwarzschild one \cite{Bergliaffa:2021diw}. 
For this static hairy Horndeski black hole, the studies of the strong gravitational lensing \cite{RKWalia,Kumar:2021cyl,Atamurotov:2022slw} and the black hole image \cite{Shi:2024bpm,Wang:2023vcv,Ghosh:2022kit,Vagnozzi:2022moj,Afrin:2021wlj} show that the Horndeski hair has a significant influence on the related observable.  Moreover, a rotating black hole with the Horndeski hair has been obtained recently by making use of the revised Newman-Janis algorithm \cite{RKWalia}. The imprints of Horndeski hair together with black hole spin  have been extensively studied in various theoretical and observational phenomena. For example, the hairy parameter  makes the shadow of a black hole to have a larger size and a greater degree of deformation.  The presence of hair in the black hole enhances the superradiance and its frequency range \cite{Jha:2022tdl}, so the energy extraction within the Horndeski framework is more efficient than in the pure Kerr case. Furthermore, the effects of the hairy parameter on the black hole stability  under  perturbations \cite{Lei:2023wlt,Yang:2023lcm}, accretion process \cite{Donmez:2024lfi}, and precessing and periodic orbits \cite{Lin:2023eyd,Lin:2023rmo} have been  studied in this rotating hairy black hole spacetime. These novel properties induced by the hairy parameter are helpful for understanding the gravitational structure of the specific theory.

In this work, we will study the near-horizon polarized images of the rotating hairy black hole in the Horndeski theory. Our study motivates from two aspects. One is to check whether the observed EVPA of electromagnetic emission from the near-horizon region is independent of the plasma flows and is determined only by the spacetime geometry in the rotating hairy Horndeski black hole spacetime. The other is to probe how the hairy parameter affects the observed EVPA in the near-horizon images, which is crucial to the issue that whether the near-horizon polarized images provide a powerful way of detecting black hole hair, as well as examining the theories of gravity.

This work is organized as follows.  In Sec. \ref{dddd}, we present the observed EVPA for the electromagnetic emission from the near-horizon region of the rotating hairy Horndeski black hole through analyzing the equatorial plasma flows with the ideal conductivity as in \cite{Hou:2024qqo}. In Sec. \ref{hhhha}, we analyze the effects of the hairy parameter on the observed EVPA in the near-horizon images, and further compare with the effects from the black hole spin and observation inclination. Finally, we end the work with a summary.

\section{Observed polarization near the horizon of a rotating hairy Horndeski black hole}
\label{dddd}

\begin{figure}[ht!]
	\includegraphics[width=3.0in]{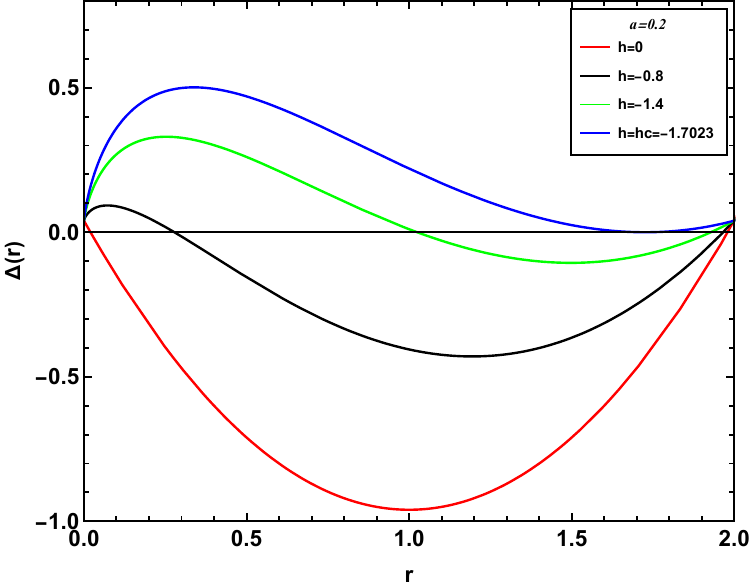} \includegraphics[width=3.0in]{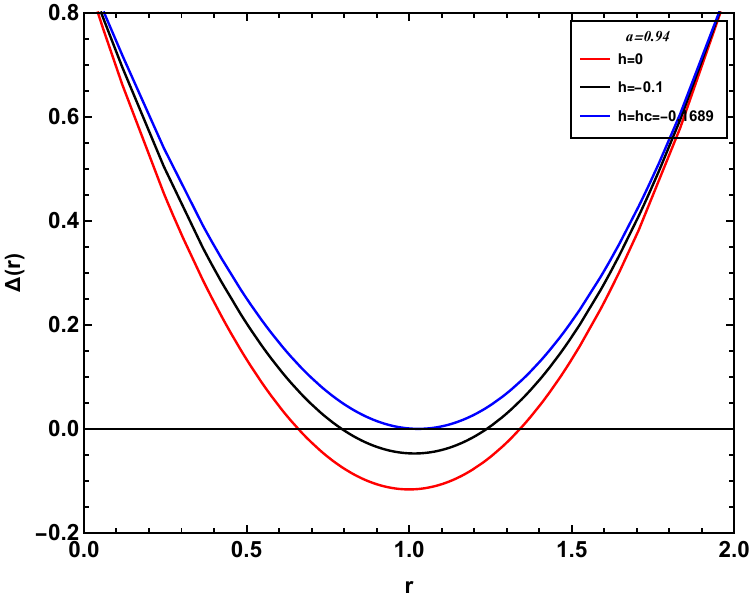}
\caption{Variation of the function $\Delta{(r)}$ with the radial coordinate $r$ for different black hole hair parameters. The black hole spins in the left and right panels are set to $ a=0.2$ and $a=0.94$, respectively. Here set $M=1$.}
	\label{images}
\end{figure}

\begin{figure}[ht!]
	\includegraphics[width=3.0in]{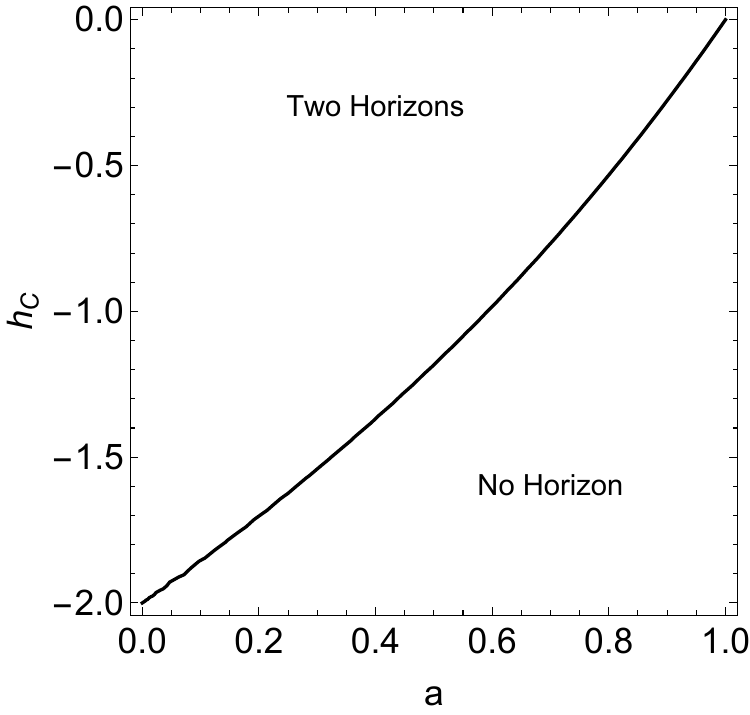} 
\caption{Variation of the critical value $h_c$ with the spin $a$ for rotating hair black holes within the Horndeski gravity. Here set $M=1$.}
	\label{criticalhc} 
\end{figure}

In this section, we will investigate the observed polarization in the near-horizon  images of a rotating hair black hole within the Horndeski gravity. Let us now to review briefly the rotating hairy Horndeski black hole, which is a stationary and spherically symmetric black hole solution with the scalar hair. With the standard Boyer-Lindquist coordinates, its metric has a form \cite{RKWalia}
\begin{eqnarray}
ds^2=&-&\left(\frac{\Delta-a^2\sin^2\theta}{\Sigma}\right)dt^2+\frac{\Sigma}{\Delta}{dr^2}+\Sigma d\theta^2+\frac{2a\sin^2\theta}{\Sigma}\left[\Delta-(r^2+a^2)\right]dt d\varphi\nonumber\\
&+&\frac{\sin^2\theta}{\Sigma}\left[(r^2+a^2)^2-\Delta a^2\sin^2\theta\right]d\varphi^2,
\label{h7}
\end{eqnarray}
with
\begin{equation}
\Delta=r^2+a^2-2Mr+hr\ln\left(\frac{r}{2M}\right), \quad\quad\quad \Sigma=r^2+a^2\cos^2\theta.
\label{h8}
\end{equation}
Here  $M$ and $a$ respectively correspond to the mass and spin parameters of the black hole, and the parameter $h$ is the hair parameter with length dimension. The metric (\ref{h7}) reduces to the Kerr case as the parameter $h\rightarrow0$, and to the static hairy Horndeski black hole \cite{Bergliaffa:2021diw} as $a=0$.
From the metric (\ref{h7}), one can find that the black hole horizons are related to the parameter $h$. Fig. \ref{images} shows the relation between the black hole horizon and the hair parameter $h$ for different black hole spins. We can find that for $h_c<h\leq0$, the black hole has  two horizons: the outer horizon $r_+$ and the inner horizon $r_-$. When $h=h_c$, the two horizons coincide and the black hole becomes an extremal one. Moreover, the critical value $h_c$ increases monotonically with the spin of the black hole, which is also shown in Fig. \ref{criticalhc}.

We are now in position to study the polarization vector of the plasma near the horizon of hairy Horndeski black holes.  As a main component in the accretion disk around black holes, the plasma is a state of matter composed of charged particles, including electrons and ions. Generally, the dynamics of plasma near black holes can be described by the general relativistic magnetohydrodynamics (GRMHD). In Ref. \cite{Hou:2023bep}, a simplified magnetohydrodynamic model has been constructed  within the GRMHD framework and the corresponding magnetic field of flows is expressed as \cite{Ruffini:1975ne,Hou:2023bep}
\begin{eqnarray}\label{fbmu}
B^{\mu} = \frac{\Psi}{\sqrt{-g}U^r}( U_{t} U^{\mu}+\delta_{t}^{\mu} ),
\label{bmu}
\end{eqnarray}
where $g$ is the determinant of the spacetime metric and $U^{\mu}$ is the plasma bulk velocity. The quantity $\Psi$ denotes the overall strength and is conserved along the bulk velocity. Since the millimeter-wavelength emission is dominated by the collective synchrotron radiation of thermal or non-thermal electrons, its polarization vector $f^{\mu}$ is largely perpendicular to the global magnetic field \cite{1979Lightman}. Therefore, the  polarization vector $f^{\mu}$ of the plasma can be written as in a covariant form \cite{Hou:2024qqo}
\begin{eqnarray}
f^\mu = \frac{\epsilon^{\mu\nu\alpha\beta} U_\nu p_\alpha B_\beta }{\omega\sqrt{B_{\bot}^2} }\,, 
\label{fmu}
\end{eqnarray}
where  $\varepsilon_{\mu\nu\alpha\beta}$ denotes the
Levi-Civita tensor and $p^{\mu}$ represents the photon four-momentum. $\omega=-U^{\mu}p_{\mu}$ is the photon frequency measured in the plasma frame. $B^{\mu}_{\bot}$ represents the spatial magnetic field in the plasma frame with the form
\begin{eqnarray}
B^{\mu}_{\bot}=B^{\mu}-\omega^{-2}(B\cdot p) p^{\mu}_{\bot}, 
\label{bcu}
\end{eqnarray}
where $p^{\mu}_{\bot}$ is the spatial four-momentum
of the photon 
\begin{eqnarray}
p^{\mu}_{\bot}=p^{\mu}-\omega U^{\mu}.
\label{bcup}
\end{eqnarray}
The celestial coordinates $(\alpha,\beta)$ of a photon emitted by the plasma synchrotron radiation in the screen of a distant observer are related to their conserved quantities  $(\lambda,\eta)$  by \cite{Himwich:2020msm,Bardeen:1973tla}
\begin{eqnarray}
\alpha=-\frac{\lambda}{\sin\theta_{0}},\quad\quad\quad \beta=\pm_{o}\sqrt{\eta+a^2\cos^2\theta_{0}-\lambda^2\cot^2\theta_{o}},
\label{ap}
 \end{eqnarray}
where $\theta_{o}$ represents the observational angle and $\pm_{o}$ denotes the sign of $p_{\theta}$ at the observer. With the components
of $f^{\mu}$ projected along the $\alpha$ and $\beta$ directions on the observer’s screen, one can obtain the EVPA 
\begin{eqnarray}\label{eevpa}
&&\chi =  \tan^{-1}\left({ \frac{f^{\beta}}{f^{\alpha}} }\right)\,.
\end{eqnarray}
 Since the geometry of a rotating hairy Horndeski black hole (\ref{h7}) belongs to a family of Petrov Type $D$ spacetimes, the Penrose-Walker quantity $\kappa$ \cite{Walker:1970un, Chandrasekhar:1985kt}
\begin{eqnarray} \label{kappa}
\kappa = \kappa_1 + i \kappa_2 = 2 \Psi_2^{-\frac{1}{3}} p^{\mu} f^{\nu} \left( l_{[\mu}n_{\nu]} - m_{[\mu}\bar{m}_{\nu]} \right) \, ,
\end{eqnarray}
is conversed along null geodesics. $l_{\mu}$, $n_{\mu}$, $m_{\mu}$ and $\bar{m}_{\mu}$ are vectors in the null basis $(l,n,m,\bar{m})$. With the four-momentum of the photon $p^{\mu}$ and the polarization vector $f^{\mu}$, the Penrose-Walker constant in the rotating hairy Horndeski black hole spacetime (\ref{h7}) can be expressed as
\begin{eqnarray}  \label{PW11}
\kappa = \kappa_{1}+i \kappa_{2}=(A-iB)\Psi_2^{-\frac{1}{3}}, 
\end{eqnarray}
with
\begin{eqnarray}  
&&A = 2(p^{[t} f^{r]}+a\sin^2\theta p^{[t} f^{\phi]}),\\ \label{PW12}
&&B = 2\sin{\theta}\left[(r^2+a^2) p^{[\phi} f^{\theta]} -a p^{[t} f^{\phi]}\right]\, ,
\end{eqnarray}
where $\kappa_1$ and $\kappa_2$ represent the real and imaginary parts of $\kappa$. The non-vanishing Weyl scalar $\Psi_2$ has a form
\begin{eqnarray}
&&\Psi_2= -\frac{1}{2(r - i a \cos\theta)^2}\bigg[\frac{2M-h}{(r-ia\cos\theta)}+\frac{h(5r+ia\cos\theta)}{6r(r+ia\cos\theta)}\bigg].
 \label{PW12}
\end{eqnarray}
 As in Ref. \cite{Hou:2024qqo}, to analyze the near-horizon features
of the polarizations of emissions from the plasma on the equatorial
plane, one can assume that the flow's velocity component $U^{\theta} = 0$, while the covariant components $U_t$ and $U_{\phi}$ keep invariant near the black hole horizon. The main reason is that the accretion lies on the equatorial plane and the near-horizon
plasma flow is dominated by the gravitational attraction, which means that the $U_t$ and $U_{\phi}$ are finite and change slowly in the near-horizon region. This assumption is supported by the simulated results of MADs presented in \cite{Chael:2021rjo,Ricarte:2022wpd}. With the above assumption, we find that
the four-velocity of the equatorial accreted flow near the outer horizon of a rotating hairy Horndeski black hole (\ref{h7}) can be expressed as
\begin{eqnarray}\label{eou}
U^t =&-& \frac{4(1-\frac{h}{2}\ln{\frac{r_+}{2}})^2(U_t+\Omega_hU_{\phi})}{\Delta }-\bigg[1+\frac{2}{r_+}\bigg(1+h-\frac{h}{2}\ln\frac{r_+}{2}-\frac{h(2r_++h)}{2(r_+-1+\frac{h}{2}+\frac{h}{2}\ln\frac{r_+}{2})}\bigg)\bigg]U_t\nonumber\\
&+&\frac{aU_{\phi}(1+\frac{h}{2}-\frac{h}{2}\ln{\frac{r_+}{2}})}{r_+^2(r_+-1+\frac{h}{2}+\frac{h}{2}\ln\frac{r_+}{2})}+ o(\Delta^0),  \\
U^r&=& \frac{-2|(1-\frac{h}{2}\ln{\frac{r_+}{2}})(U_t+\Omega_hU_{\phi})|}{r_+} + o(\Delta^{0}),\label{eou1}\\
U^{\theta}& =& 0 ,\\
U^{\phi} &=& -\frac{2a(1-\frac{h}{2}\ln{\frac{r_+}{2}})(U_t+\Omega_hU_{\phi})}{\Delta r_+} + \frac{(1+\frac{h}{2}-\frac{h}{2}\ln\frac{r_+}{2})(U_{\phi}+aU_t)}{r_+^2(r_+-1+\frac{h}{2}+\frac{h}{2}\ln\frac{r_+}{2})}+ o(\Delta^0) \, \label{eou2},
\end{eqnarray}
where $\Omega_h=\frac{a}{r^2_++a^2}$ is the angular velocity at the outer horizon. Obviously, as $h\rightarrow 0$, the four-velocity of the equatorial accreted flow $U^{\mu}$ tends to that in the Kerr case obtained in Ref. \cite{Hou:2024qqo}. At the outer horizon, from $\Delta(r_+)=0$, one has $r^2_++a^2=2r_+(1-\frac{h}{2}\ln\frac{r_+}{2})>0$, which means that $U^r<0$ for the accreting plasma as in the Kerr black hole case. Inserting Eqs. 
 (\ref{eou})-(\ref{eou2}) into Eq. (\ref{fbmu}), one can obtain the near-horizon expansion of the magnetic field
\begin{eqnarray}\label{eob}
&&B^t = \frac{2\Psi U_t(1-\frac{h}{2}\ln\frac{r_+}{2})}{r_+\Delta} \frac{(U_t+\Omega_hU_\phi)}{|U_t+\Omega_hU_\phi|} + o(\Delta^{-1}), \\
&&B^r = \frac{\Psi}{r_+^2} U_t, \\
&&B^{\theta} = 0,\\
&&B^{\phi} = \frac{\Psi U_t a}{r_+^2 \Delta} \frac{(U_t+\Omega_hU_\phi)}{|U_t+\Omega_hU_\phi|} + o(\Delta^{-1}) \, .
\end{eqnarray}
Moreover, for the plunging matter, one has $U_t+\Omega_hU_{\phi}< 0$ since a probe particle with $U_t+\Omega_hU_{\phi}>0$ cannot classically reach the event horizon. For the roating hairy Horndeski black hole, we find that the ratio also meets $B^{\phi}/B^r\approx -a/ \Delta$, which agrees with that from the simulation presented in \cite{Ricarte:2022wpd}. Expanding
the polarization vector $f^{\mu}$ near the outer horizon $r = r_+$, one has
\begin{eqnarray}\label{eof}
&& \frac{X}{\varepsilon} f^{t} = \frac{-a(a^2+r_+^2)\sigma_{\theta}\sqrt{\eta}}{ r_+^2\Delta } + \frac{a\sigma_{\theta}\sqrt{\eta}}{r_+^2}+\frac{2a^3\sigma_{\theta}\sqrt{\eta}}{r_+^3(h+2r_++h\ln{\frac{r_+}{2}})} +o(\Delta^0) \\
&&\frac{X}{\varepsilon} f^{r} =  \frac{a\sigma_{\theta}\sqrt{\eta}}{ r_+^2} +R_1 \Delta+o(\Delta) \\
&& \frac{X}{\varepsilon} f^{\theta} = \frac{a(a\lambda-a^2-r_+^2)-\sigma_r a |a\lambda-a^2-r_+^2|}{r_+^2\Delta} + \Theta_1 + o(\Delta^0)   \\\label{eof2}
&&\frac{X}{\varepsilon} f^{\phi} =  -\frac{a^2\sigma_{\theta} \sqrt{\eta}}{ r_+^2\Delta } + \frac{\sigma_{\theta}\sqrt{\eta}}{r_+^2} \frac{2a^3\sigma_{\theta}\sqrt{\eta}}{r_+^3(h+2r_++h\ln{\frac{r_+}{2}})} +o(\Delta^0)\,,
\end{eqnarray}
with $X =-\Psi^{-1}\sqrt{-g(U^{\mu}p_{\mu})B^2_{\bot}}$ and $\varepsilon=-p_t$. $\lambda$ and $\eta$ are conserved impact parameters for the emitted photons. The terms $\sigma_{r}$ and $\sigma_{\theta}$ denote the sign of  $p^{r}$ and $p^{\theta}$, respectively. The forms of $R_1$ and $\Theta_1$ are rather complicated and we do not list them here. 

Substituting Eqs. 
 (\ref{eof})-(\ref{eof2}) into Eq. (\ref{kappa}), we can obtain the expansion of the Penrose-Walker constant $\kappa$ near the outer horizon of the rotating hairy Horndeski black hole, and find that the ratio 
\begin{eqnarray}\label{master}
z = \frac{\kappa_1}{\kappa_2} = z_0 + z_1 \Delta + o(\Delta^2),
\end{eqnarray}
is independent of the plasma motion near the horizon as in the Kerr case \cite{Hou:2024qqo}, where the coefficients $z_0$ and $z_1$ can be expressed respectively as
\begin{eqnarray}\label{master1}
z_0 = \frac{\sigma_\theta \sqrt{\eta}}{\lambda - a}, \quad \quad \quad z_1 = \frac{\sigma_\theta\sqrt{\eta}(1+z^2_0)\Omega_h}{2a^2(\Omega_h\lambda-1)}.
\end{eqnarray}
Obviously, the forms of the coefficients $z_0$ and $z_1$ coincide with those in the Kerr case. However, we find that in the rotating hairy Horndeski black hole case, for the fixed $\lambda$ and $\eta$, the coefficient $z_0$ depends on only the black hole spin parameter $a$, while the coefficient $z_1$ is related to the hairy parameter $h$ by the angular velocity $\Omega_h$ of the outer horizon. 
According to Ref. \cite{Himwich:2020msm}, the EVPA  on the observer's plane can be derived from Eq. (\ref{eevpa}) and Eq.  (\ref{master}), which is given by
\begin{eqnarray}\label{evpa}
\chi = \tan^{-1}\left(\frac{\beta z + \mu}{\beta - \mu z}\right) =\tan^{-1}\left(\frac{\mu}{\beta}\right)+ \tan^{-1}z = \tan^{-1}\left(\frac{\mu}{\beta}\right) + \tan^{-1}z_0 + \frac{z_1 \Delta}{1 + z_0^2} + o(\Delta^1),
\end{eqnarray}
 where $\mu = -(\alpha+a \sin{\theta_o})$. 
From Eqs. (\ref{master}), (\ref{master1}) and (\ref{evpa}), it is found that the hairy parameter $h$ appears explicitly  only in the sub-leading term. However, the effects of $h$ on the observed EVPA $\chi$ also include the contributions from the leading terms because the parameters $\lambda$ and $\eta$ of photons traveling from the near-horizon region to  a specific observer are not arbitrary, but should be certain specific values related to 
the hairy parameter $h$. This means that the polarization structures near the horizon could provide a potential tool to detect black hole's hairs.
\begin{figure}[ht!]
	\centering
	\includegraphics[width=3in]{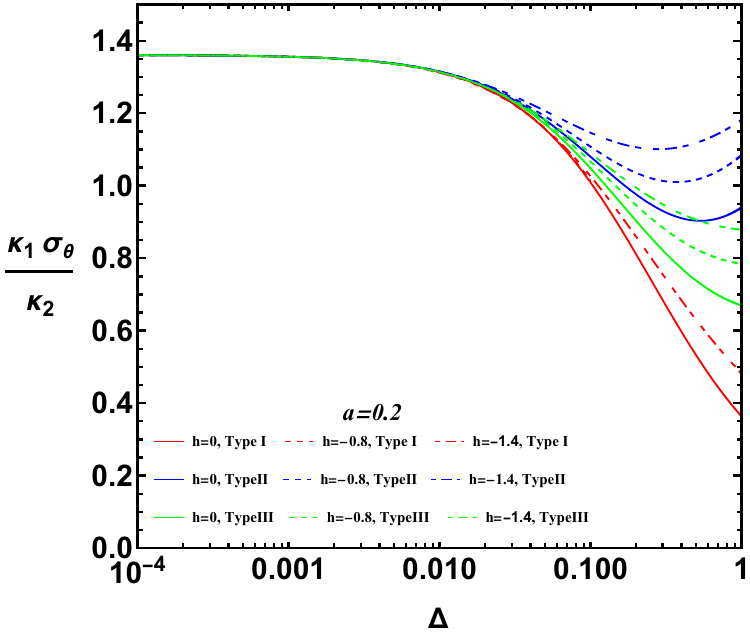}\quad
	\includegraphics[width=2.9in]{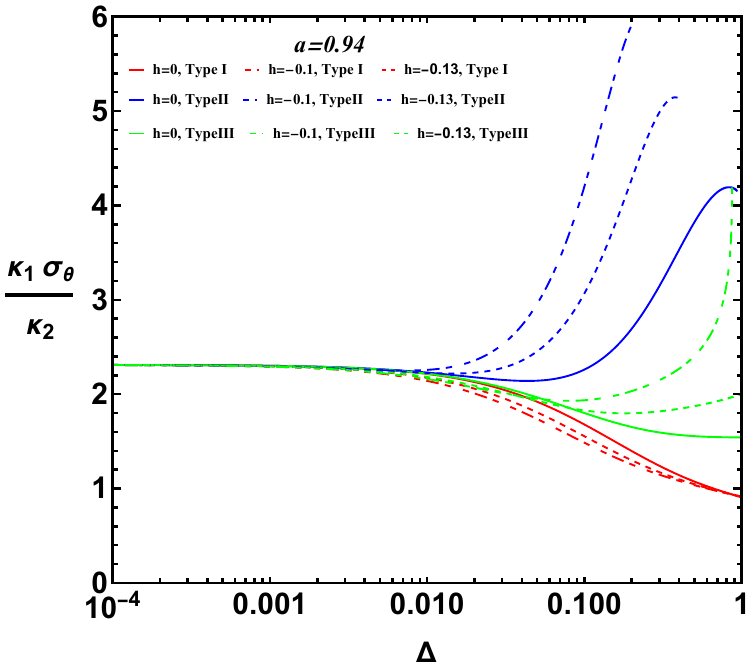}\\
        \includegraphics[width=3in]{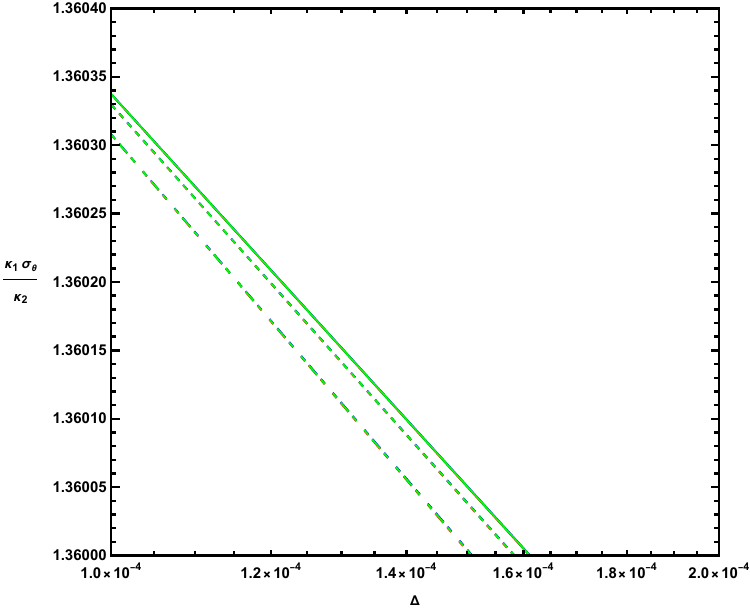}\quad
	\includegraphics[width=3.1in]{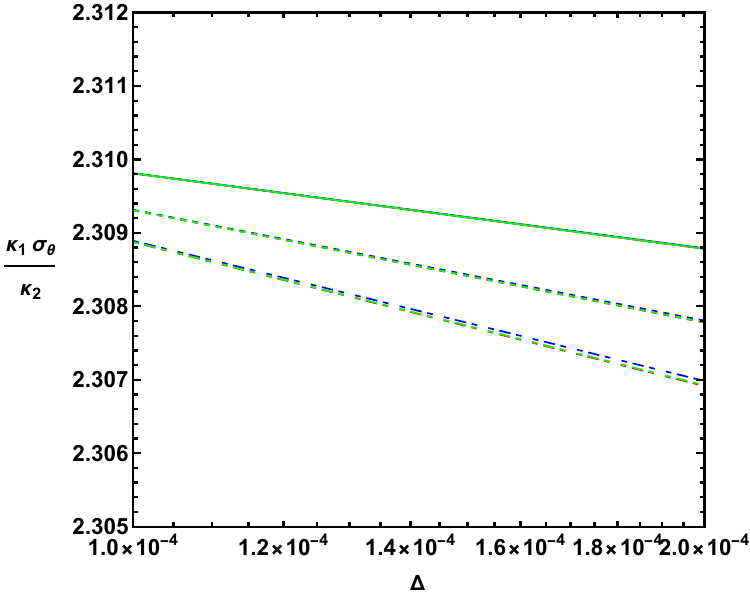}
	\centering
	\caption{Variation of  $z$ with $\Delta$ for three different types of plasma flows in the rotating hairy Horndeski black hole spacetime. The left and right panels correspond to spin parameters $a=0.2$ and $a=0.94$, respectively. Each panel in the bottom row is a partial enlargement of the respective panel in the upper row. }
	\label{test}
\end{figure}

\section{ Impacts of the Hairy Parameter on the Near-Horizon Polarized Image of a rotating Horndeski Black Hole}
\label{hhhha}

Let us now to discuss the features of the near-horizon polarized image of a rotating Horndeski black hole (\ref{h7}) and to probe the corresponding impacts of the hairy parameter $h$. In Fig. \ref{test}, we present the variations of the ratio $z$ with the quantity $\Delta$ in the rotating hairy Horndeski black hole spacetime for three different plasma flow conditions as in \cite{Hou:2023bep}. The plasma is assumed to infall from infinity in the type I model and to plunge from the prograde innermost stable circular orbit (ISCO) in the type II model, respectively. The type III model corresponds to the fitting result from the simulation of MAD introduced in \cite{Chael:2021rjo}. 
Our results show that the curves  $z-\Delta$ are almost overlapped in the range $\Delta<10^{-2}$ for all three types of plasma flows in the black hole spacetime with different values of the hairy parameter $h$. This further confirms that the effects arising from types of plasma flows on the polarization structures near the black hole horizon are negligible. Furthermore, from their partial enlargements, 
we find that the curves  $z-\Delta$ are separated for different hairy parameters $h$, while still are overlapped for above three types of plasma flows. This means that the effects from the hairy parameter $h$ is much stronger than those from the types of plasma flows.  In addition,  the ratio $z$ increases with the increase of the hairy parameter $h$. 

\begin{figure}[ht!]
\includegraphics[width=3in]{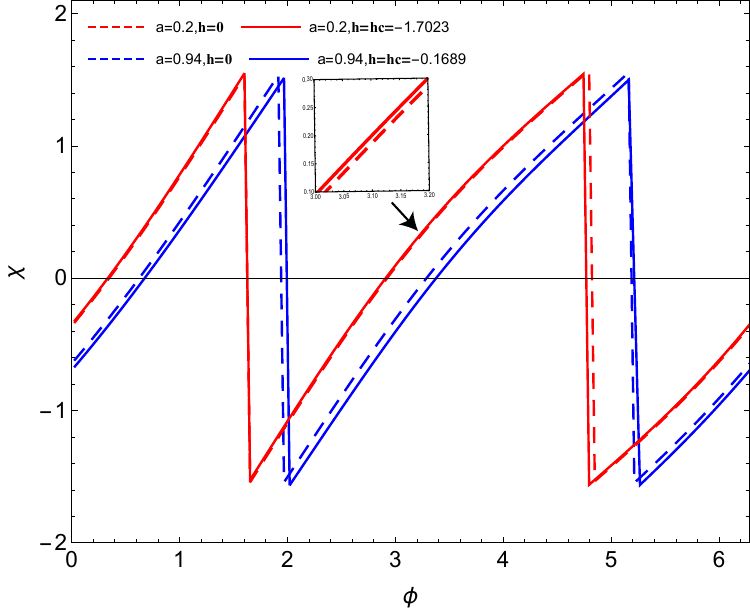}\quad\includegraphics[width=3in]{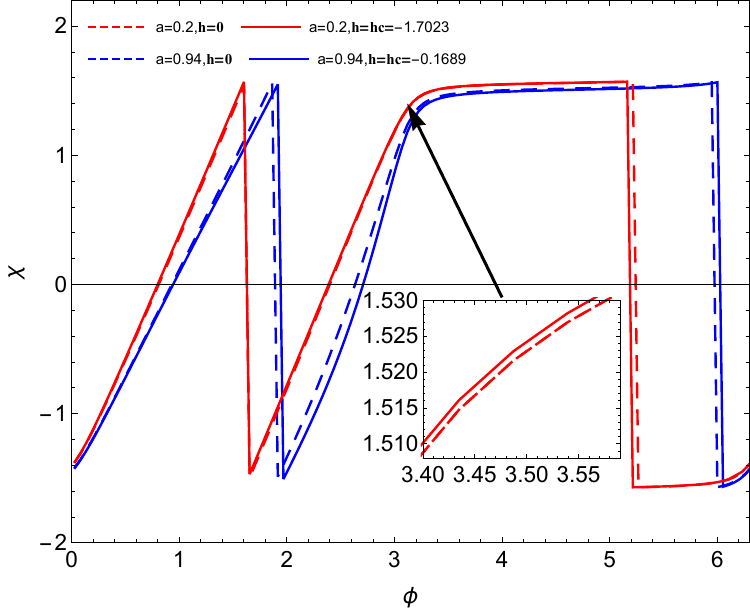}
	\caption{Variation of the EVPA in the near-horizon images with the azimuthal angle $\phi$  in the rotating hairy Horndeski  black hole spacetime. The left panel is for $\theta_o=17^{\circ}$ and the right one is for $\theta=80^{\circ}$. In each panel, the solid and dashed lines correspond to the cases $h=h_c$ and $h=0$, and the blue and red lines denote the cases $a=0.94$ and $a=0.2$, respectively.  }
	\label{NHI01}
\end{figure}

In Fig. \ref{NHI01}, we present the changes of the observed EVPA of polarization vectors with the azimuth angle $\phi$ for the photons from region near the event horizon for a rotating hairy Horndeski black hole with different hairy parameters $h$.
As expected, the observed EVPA of polarization vectors significantly changes with the black hole's spin $a$ and the observed inclination $\theta_o$. Furthermore, we find that the effects of hairy parameter $h$ on the observed EVPA $\chi$ depend on the black hole's spin $a$ and the observed inclination $\theta_o$. For a fixed inclination $\theta_o$, in the slowly rotating case where $a=0.2$, the observed EVPA 
$\chi$ corresponding to $h=h_c$ is greater than that associated with $h=0$.
The situation is reversed in the rapidly rotating case with $a=0.94$. For a fixed black hole spin $a$, as the  azimuthal angle $\phi$ is in the range $(3,6)$, one can find that the impact of $h$ on the observed EVPA is more distinct in the low inclination case with $\theta_o=17^{\circ}$. As the azimuthal angle $\phi$ is beyond the range $(3,6)$,  the difference between the observed EVPAs induced by the hairy parameter $h$ is almost independent of the observer inclination $\theta_o$. Therefore,  the observer inclination's impact on the EVPA difference caused by the hairy parameter $h$ is also shaped by the azimuthal angle of the observed polarization vector. Through making a comparison, we also find that the dependence of the hairy parameter's impact on the EVPA is stronger with respect to the black hole spin $a$ than to the inclination angle $\theta_o$.

\begin{figure}[ht!]
\includegraphics[width=3in]{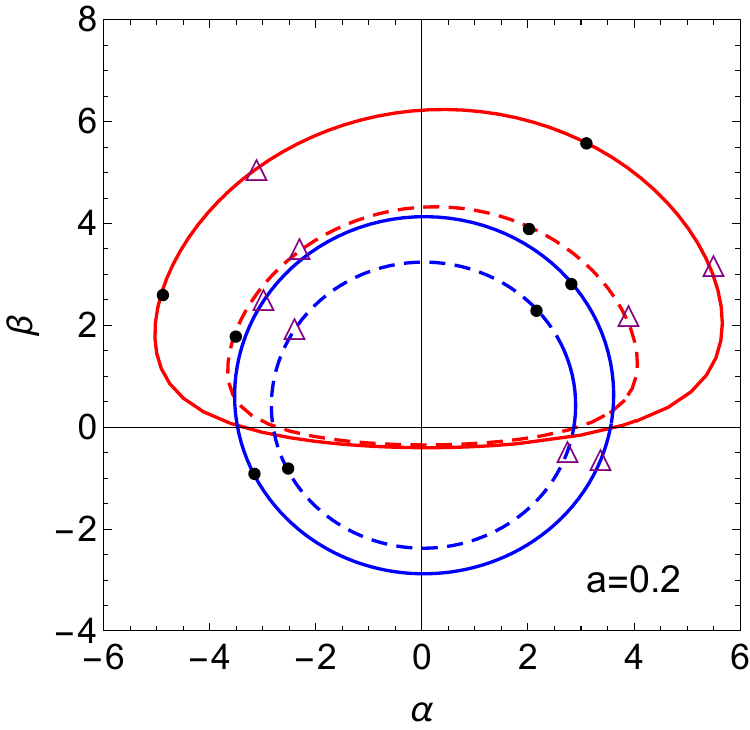}\quad\includegraphics[width=3in]{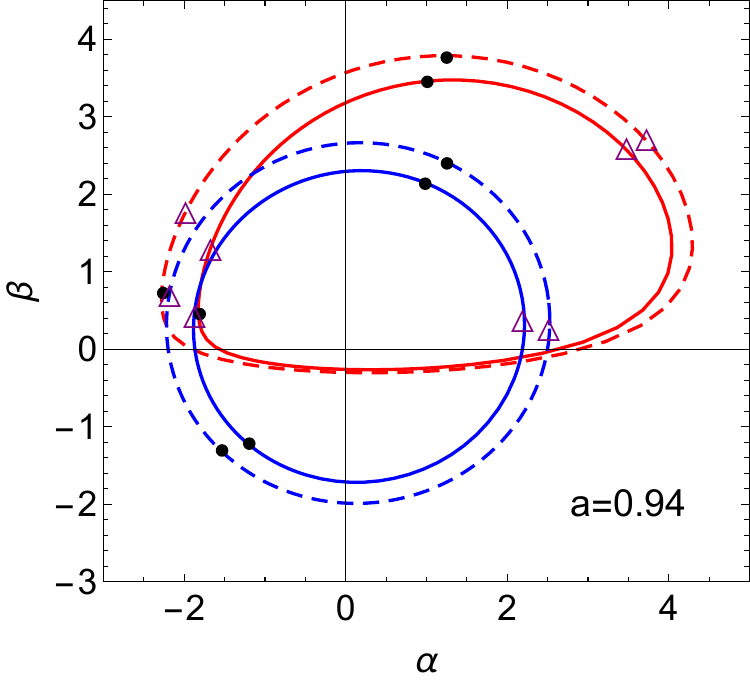}
	\caption{EVPA in the near-horizon images  of rotating hairy Horndeski black holes with different symbols: $\bullet$ indicates $\chi=\frac{\pi}{6}$, and $\triangle$ indicates $\chi=-\frac{\pi}{6}$. The left panel is for $a=0.2$ and the right one is for $a=0.94$. In each panel, the curves represent the near-horizon images. The solid and dashed lines correspond to the cases  $h=h_c$ and $h=0$, and the blue and red lines denote the cases $\theta_o=17^{\circ}$ and $\theta_o=80^{\circ}$, respectively. }
	\label{NHI02}
\end{figure}

In Fig. \ref{NHI02}, we present the positions of the polarization vectors related to EVPA $\chi=\frac{\pi}{6}$ or $\chi=-\frac{\pi}{6}$ in the near-horizon images for a rotating hairy Horndeski black hole with spin parameters $a =0.2$ and $a = 0.94$, observed at inclination angles $\theta_o = 17^{\circ} $ and $\theta_o = 80^{\circ}$.
It is shown that the position of polarization vectors with the same EVPA $\chi=\frac{\pi}{6}$ or $\chi=-\frac{\pi}{6}$ changes with the hairy parameter $h$. 
Apart from the size of the event horizon image, the variation in the position of polarization vectors with the same EVPA, induced by the hairy parameter $h$, is most strongly affected by the black hole spin. Although the effects from the hairy parameter $h$ in the near-horizon images are weaker than those from the black hole spin and observed inclination, one can infer the information on the black hole's hair by integrating the corresponding observations with the higher precision.

\section{Discussions and conclusions}
\label{Conclusion}

We have investigated the observed polarization in the near-horizon  images of a rotating hair black hole within the Horndeski gravity.  Besides its mass and spin parameters, this black hole possesses an additional hairy parameter $h$, which originates from the Horndeski theory. Our results confirm once again that the EVPA of polarization vectors in the near-horizon images is governed exclusively by the spacetime geometry and is irrespective of the plasma flow geometry as found in \cite{Hou:2024qqo}. We observe that the effects of the hairy parameter $h$ on the observed EVPA  depend on the black hole's spin $a$ and the observed inclination $\theta_o$. For a fixed inclination $\theta_o$ in the slowly rotating case, the observed EVPA  corresponding to $h=h_c$ is greater than that associated with $h=0$.
The situation is reversed in the rapidly rotating case.  For a fixed black hole spin, the dependence of the difference between the observed EVPAs induced by the hairy parameter $h$ on the observer inclination $\theta_o$ is also influenced by the azimuthal angle of the observed polarization vector. As the azimuthal angle $\phi$ is in a specific range, for example, the range $(3,6)$ considered in this work,  the impact of $h$ on the observed EVPA is more distinct in the low inclination case. However, as the azimuthal angle $\phi$ is beyond the specific range,  the difference between the observed EVPAs induced by the hairy parameter $h$ is almost independent of the observer inclination $\theta_o$. Furthermore, the dependence of the hairy parameter's impact on the EVPA is stronger with respect to the black hole spin $a$ than to the inclination angle $\theta_o$. Although the effects from the hairy parameter $h$ in the near-horizon images is weaker than those from black hole spin and the observed inclination, we can infer the information on the black hole's hair by integrating the corresponding observations with the higher precision, which could help to further understand the no-hair theorem.

\section*{Acknowledgements}
This work was supported by the National Natural Science Foundation of China (Grant Nos. 12275079 and 12035005), the National Key Research and Development Program of China (Grant No. 2020YFC2201400), and the innovative research group of Hunan Province under Grant No. 2024JJ1006.

\end{document}